\documentclass[letterpaper, 10 pt, conference]{ieeeconf}                               
\IEEEoverridecommandlockouts

\usepackage{array}
\newcommand{\PreserveBackslash}[1]{\let\temp=\\#1\let\\=\temp}
\newcolumntype{C}[1]{>{\PreserveBackslash\centering}p{#1}}
\newcolumntype{R}[1]{>{\PreserveBackslash\raggedleft}p{#1}}
\newcolumntype{L}[1]{>{\PreserveBackslash\raggedright}p{#1}}
\usepackage{supertabular,booktabs}
\usepackage{graphicx}
\usepackage{multirow}

\makeatletter
\def\thickhline{%
  \noalign{\ifnum0=`}\fi\hrule \@height \thickarrayrulewidth \futurelet
   \reserved@a\@xthickhline}
\def\@xthickhline{\ifx\reserved@a\thickhline
               \vskip\doublerulesep
               \vskip-\thickarrayrulewidth
             \fi
      \ifnum0=`{\fi}}
\makeatother

\newlength{\thickarrayrulewidth}
\usepackage[bottom]{footmisc} 
\usepackage{ragged2e}
\usepackage{float}

\usepackage[bottom]{footmisc} 
\usepackage[mode=buildnew]{standalone} 
\usepackage{ragged2e}
\usepackage{float}
\graphicspath{{./fig/}}
\usepackage{tabularx}
\usepackage{parskip}
\usepackage[ruled, vlined]{algorithm2e}
\usepackage{rotating} 
\usepackage{makecell}
\usepackage{enumerate}
\usepackage{longtable}
\usepackage{lettrine}
\usepackage{textcomp}
\usepackage[dvipsnames]{xcolor}
\RequirePackage[hidelinks, bookmarks, bookmarksopen=true, plainpages=false, pdfpagelabels, pdfpagelayout=SinglePage]{hyperref}
\hypersetup{
colorlinks=true,
linkcolor=Blue,
filecolor=black,      
urlcolor=black,
citecolor = red,
}
\let\oldciteauthor=\citeauthor
\def\citeauthor#1{\hypersetup{citecolor=black}\oldciteauthor{#1}}
\let\oldcite=\cite
\def\cite#1{\hypersetup{citecolor=black}\oldcite{#1}}

\def\BibTeX{{\rm B\kern-.05em{\sc i\kern-.025em b}\kern-.08em
    T\kern-.1667em\lower.7ex\hbox{E}\kern-.125emX}}

\makeatletter
\let\old@makecaption=\@makecaption
\usepackage{subcaption}
\let\@makecaption=\old@makecaption
\makeatother

\usepackage[fleqn]{amsmath}
\usepackage{amssymb,amsfonts, mathtools}
\usepackage{bbm}
\usepackage{bm}
\usepackage{cuted}
\usepackage{etoolbox}
\usepackage{upgreek}
\AtBeginEnvironment{gather}{\setcounter{equation}{0}}
\RestyleAlgo{ruled}
\SetKw{KwBy}{by}
\SetKw{KwRequire}{Require}
\SetKw{KwInput}{Input}
\SetKw{KwInputs}{Inputs}
\SetKw{KwOutput}{Output}
\SetKw{KwOutputs}{Outputs}

\newcommand{\tee}[1]{\texttt{#1}}

\newcommand{\mb}[1]{\mathbb{#1}}
\newcommand{\mbf}[1]{\mathbf{#1}}
\newcommand{\mc}[1]{\mathcal{#1}}
\newcommand{\ith}{i^{\text{th}}}

\newcommand{\x}{\mathbf{x}}

\makeatletter
\newcommand{\mathleft}{\@fleqntrue\@mathmargin0pt}
\newcommand{\mathcenter}{\@fleqnfalse}
\makeatother

\usepackage{pgf, tikz}
\usetikzlibrary{shapes,arrows,external,calc, automata, arrows.meta,patterns} 
\usepackage{pgfplots}
\pgfplotsset{width=10cm,compat=1.9}
\tikzstyle{block} = [draw, fill=green!20, text centered, rectangle,
minimum height=0.8cm,minimum width=6em, align=right]
\tikzstyle{ipe stylesheet} = [
          ipe import,
          even odd rule,
          line join=round,
          line cap=butt,
          ipe pen normal/.style={line width=0.4},
          ipe pen heavier/.style={line width=0.8},
          ipe pen fat/.style={line width=1.2},
          ipe pen ultrafat/.style={line width=2},
          ipe pen normal,
          ipe mark normal/.style={ipe mark scale=3},
          ipe mark large/.style={ipe mark scale=5},
          ipe mark small/.style={ipe mark scale=2},
          ipe mark tiny/.style={ipe mark scale=1.1},
          ipe mark normal,
          /pgf/arrow keys/.cd,
          ipe arrow normal/.style={scale=7},
          ipe arrow large/.style={scale=10},
          ipe arrow small/.style={scale=5},
          ipe arrow tiny/.style={scale=3},
          ipe arrow normal,
          /tikz/.cd,
          ipe arrows, 
          <->/.tip = ipe normal,
          ipe dash normal/.style={dash pattern=},
          ipe dash dotted/.style={dash pattern=on 1bp off 3bp},
          ipe dash dashed/.style={dash pattern=on 4bp off 4bp},
          ipe dash dash dotted/.style={dash pattern=on 4bp off 2bp on 1bp off 2bp},
          ipe dash dash dot dotted/.style={dash pattern=on 4bp off 2bp on 1bp off 2bp on 1bp off 2bp},
          ipe dash normal,
          ipe node/.append style={font=\normalsize},
          ipe stretch normal/.style={ipe node stretch=1},
          ipe stretch normal,
          ipe opacity 10/.style={opacity=0.1},
          ipe opacity 30/.style={opacity=0.3},
          ipe opacity 50/.style={opacity=0.5},
          ipe opacity 75/.style={opacity=0.75},
          ipe opacity opaque/.style={opacity=1},
          ipe opacity opaque,
        ]
\definecolor{red}{rgb}{1,0,0}
\definecolor{blue}{rgb}{0,0,1}
\definecolor{green}{rgb}{0,1,0}
\definecolor{yellow}{rgb}{1,1,0}
\definecolor{orange}{rgb}{1,0.647,0}
\definecolor{gold}{rgb}{1,0.843,0}
\definecolor{purple}{rgb}{0.627,0.125,0.941}
\definecolor{gray}{rgb}{0.745,0.745,0.745}
\definecolor{brown}{rgb}{0.647,0.165,0.165}
\definecolor{navy}{rgb}{0,0,0.502}
\definecolor{pink}{rgb}{1,0.753,0.796}
\definecolor{seagreen}{rgb}{0.18,0.545,0.341}
\definecolor{turquoise}{rgb}{0.251,0.878,0.816}
\definecolor{violet}{rgb}{0.933,0.51,0.933}
\definecolor{darkblue}{rgb}{0,0,0.545}
\definecolor{darkcyan}{rgb}{0,0.545,0.545}
\definecolor{darkgray}{rgb}{0.663,0.663,0.663}
\definecolor{darkgreen}{rgb}{0,0.392,0}
\definecolor{darkmagenta}{rgb}{0.545,0,0.545}
\definecolor{darkorange}{rgb}{1,0.549,0}
\definecolor{darkred}{rgb}{0.545,0,0}
\definecolor{lightblue}{rgb}{0.678,0.847,0.902}
\definecolor{lightcyan}{rgb}{0.878,1,1}
\definecolor{lightgray}{rgb}{0.827,0.827,0.827}
\definecolor{lightgreen}{rgb}{0.565,0.933,0.565}
\definecolor{lightyellow}{rgb}{1,1,0.878}
\definecolor{black}{rgb}{0,0,0}
\definecolor{white}{rgb}{1,1,1}

\usepackage{epsfig}

\usepackage{soul}

\usepackage{capt-of}

\definecolor{mplblue}{RGB}{31, 119, 180}

\usepgfplotslibrary{external} 
\usetikzlibrary{external}
\newcommand{\sfigsz}{0.9\columnwidth}

\begin{document}
\mathcenter

\title{Distributed Optimal Formation Control for an Uncertain Multiagent System in the Plane\\
\thanks{\textsuperscript{1}C. Enwerem and J. Baras are with the Department of Electrical \& Computer Engineering and the Institute for Systems Research, University of Maryland, College Park, MD 20742, USA.}
\thanks{
\textsuperscript{2}D. Romero is with the Department of Electrical \& Computer Engineering, University of Maryland, College Park, MD and the University System of Maryland at Southern Maryland, California, MD 20619, USA.
Emails: \{\texttt{enwerem, baras, dbromero}\}@umd.edu
}

\thanks{*This work received support in part from the Office of Naval Research (Grant No. N000141712622), the Naval Air Warfare Center Aircraft Division Seed Grant (Grant No. N004212120001), and from a joint award from Microsoft Corporation, 1 Microsoft Way, Redmond, WA 98052, USA and the Maryland Robotics Center, 3156 Brendan Iribe Center, College Park, MD 20742, USA.}
}

\author{{Clinton Enwerem\textsuperscript{1}}
\and
{John Baras\textsuperscript{1}}
\and
{Danilo Romero\textsuperscript{2}}
}

\maketitle
\thispagestyle{plain}
\pagestyle{plain}

\begin{abstract}
In this paper, we present a distributed optimal multiagent control scheme for quadrotor formation tracking under localization errors. Our control architecture is based on a leader-follower approach, where a single leader quadrotor tracks a desired trajectory while the followers maintain their relative positions in a triangular formation. We begin by modeling the quadrotors as particles in the YZ-plane evolving under dynamics with uncertain state information. Next, by formulating the formation tracking task as an optimization problem --- with a constraint-augmented Lagrangian subject to dynamic constraints --- we solve for the control law that leads to an optimal solution in the control and trajectory error cost-minimizing sense. Results from numerical simulations show that for the planar quadrotor model considered --- with uncertainty in sensor measurements modeled as Gaussian noise --- the resulting optimal control is able to drive each agent to achieve the desired global objective: leader trajectory tracking with formation maintenance. Finally, we evaluate the performance of the control law using the tracking and formation errors of the multiagent system.
\end{abstract}


\maketitle


\begin{figure*}
    \centering
    \includegraphics[width=0.7\textwidth]{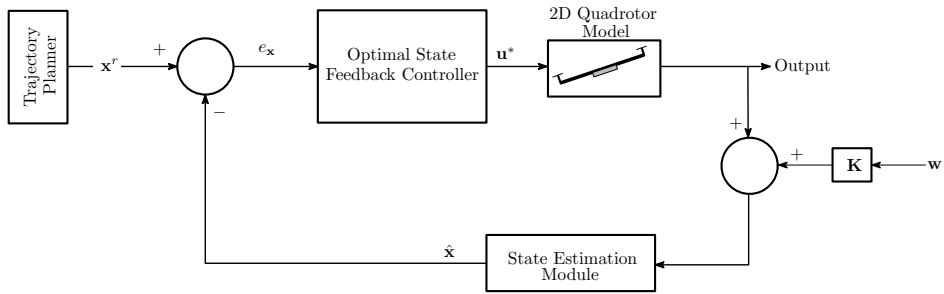}
    \caption{Block Diagram of our proposed control architecture.}
    \label{fig:optcontarch}
\end{figure*}

\section{Introduction} 
The task of formation control is central to many problems in multiagent coordination and cooperative control, as it is usually the first problem one typically has to solve to achieve some collective objective with multiple agents. In the standard formation control problem, it is usually of interest to control a group of agents --- so that they converge to unique terminal states and with the goal of attaining a desired geometric pattern --- to facilitate a specific task. Such a control objective finds direct application in several areas such as reconnaissance, aerial coverage and monitoring, mobile target tracking, and in mobile communication network maintenance, to name a handful. In problems involving formation control, the prevailing assumptions are usually that all the agents have either the same forward velocity \cite{maCooperativeTargetTracking2015a, marshallFormationsVehiclesCyclic2004} or angular velocity \cite{brinonarranzContractionControlFleet2010}, and that information about the state of each agent is available to its neighbors. The agents obtain this state information from either a central station broadcasting to all agents or from a more complex distributed network topology that can be fixed, stochastic, or even have intrinsic dynamics \cite{caoOverviewRecentProgress2013}.

Conventionally, the formation control problem takes one of two broad forms: group reference formation control and non-group reference formation control \cite{caoOverviewRecentProgress2013}. Group-reference formation control, also known as formation tracking, is the case where the agents move in formation while tracking a reference trajectory or \textit{group reference}. In non-group reference formation control on the other hand, the agents are tasked with maintaining a specific geometric shape without following any trajectory setpoint. 

Unsurprisingly, much of the research on formation control is centered around the more challenging problem of formation tracking, and several methods have been proposed (see \cite{ohSurveyMultiagentFormation2015} for a detailed survey on the topic). There is the well-researched leader-follower paradigm where one agent is taken as the leader and the other agents, as followers, that must track the leader's motion while maintaining some pre-specified distance from themselves and the leader. Defining rules that govern the evolution of these inter-agent distances thus leads to the desired formation, and by varying the rules, a new formation results. Simultaneous tracking under this formation is then achieved by specifying the desired trajectory as the leader's path setpoint.

To effectively track the leader, follower agents require sufficiently accurate estimates of the leader's pose in the inertial frame, which can be affected by noisy sensor measurements, exogenous disturbances from the environment, such as wind or downwash from nearby agents --- in the case where the agents are aerial vehicles --- or even uncertainty in the communication network from delays and packet drops. Thus, it is often the case that the multiagent system (MAS) will fail to track the reference trajectory while keeping formation, or deviate from the desired formation altogether, causing unintended and even unsafe effects \cite{liuFurtherResultsDistance2021}. Furthermore, the disturbances themselves may be difficult, computationally expensive, or impossible to estimate, making formation tracking under uncertainty both a safety-critical requirement and a nontrivial problem. 

Several studies have approached the formation tracking problem from an optimal control viewpoint. One of the earliest efforts at formulating the tracking with formation maintenance task as an optimal control problem was presented in \cite{wangIntegratedOptimalFormation2013}. Here, using an approach derived from the Riccati equation, the authors designed a distributed optimal formation control law --- for multiple UAVs with linear models --- by minimizing a non-quadratic cost function. The optimal control formulation was given here, however, without any consideration to the pairwise distances between agents. Following standard thinking based on Pontryagin's Minimum Principle (PMP), the authors in \cite{liuFinitetimeFormationControl2015} presented an optimal formation control approach by minimizing the control energy of the system, with the agents evolving under perfect-state dynamic models. More recently, an identifier-critic-actor reinforcement learning based method was employed in \cite{wenOptimizedFormationControl2020} to select the optimal control policy for an MAS comprising agents with unknown and adaptively-identified nonlinear dynamics. 

In our work, we study the problem of formation tracking under localization errors where the leader in the MAS is required to track a sinusoidal reference. Simultaneously, the followers are required to keep their assigned planar positions with respect to the leader and themselves as defined by a triangular formation rule. In contrast to the aforementioned research articles, our work focuses on designing optimal formation tracking laws for a specific case where the agents are modeled as quadrotors in the plane under uncertainty (from sensor noise). We also formulate the formation tracking task as a dynamic optimization problem with a constrained-augmented Lagrangian and solve it using optimization software tools, as opposed to traditional analytical optimal control methods like PMP or the Riccati equation.

\subsection{Contributions}
 Our contributions are as follows:
  \begin{enumerate}[i.]
      \item Application of optimal control theory to a uniquely-formulated multiagent formation tracking problem.
      \item Simulative validation of the effectiveness of the optimal control law in both the nominal setting and the case with Gaussian noise in state measurements.
 \end{enumerate}

In what follows, we introduce the notation used in this work and discuss the setting under which we study the formation tracking problem (see Section \ref{sec:probform}). Next, in Section \ref{sec:2dquad}, we provide details about the planar quadrotor model under consideration. Section \ref{sec:optconc} puts forward the optimal control component of our work. Following that, in Section \ref{ssec:formgen}, we discuss motivations for electing a triangular formation as the reference formation in our work, along with a brief description of the properties of this desired formation. The simulation setup is provided in Section \ref{sec:simstudies}, with key simulation results following in Section \ref{sec:res}. Finally, we conclude the paper with recommendations for future research in Section \ref{sec:conc}. 


\section{Preliminaries}
\label{sec:probform}
We pose the formation tracking problem, as considered in this article, under the following assumptions:
\begin{enumerate}[i.]
    \item All agents are homogeneous, i.e., they are identical, hence (\ref{it:limod}) follows.
    \item \label{it:leadfoll} All but one (randomly-chosen) agent (the leader)  belong to the follower group; information about the leader's state is available to all the follower agents through a common communication network shared by all agents.
    \item \label{it:limod} The model of each agent can be approximated by a linear time-invariant continuous-time model. See Section \ref{sec:2dquad} for a description of the model.
    \item Each agent's roll angle -- and thus, rate -- is approximately zero, i.e., the agent moves to its position in the formation by maintaining a near-hover state.
    \item Each follower agent is driven independently to execute the local task of keeping its pre-assigned position in the inertial frame and also to simultaneously achieve the collective task of maintaining a desired group formation with the other agents. This assumption implies that there are no adversarial agents within the group.
    \item The uncertainty in the MAS is only due to localization errors from the state estimation module (see Figure \ref{fig:optcontarch}), hence the agents' states are perturbed by sensor noise, and are thus taken to be imperfect. Effects from external disturbances such as wind gust and downwash are neglected.
\end{enumerate}

We denote the $i^{\text{th}}$ agent as $a_i \in \mathcal{A}$, where $\mathcal{A}$ is the set of all agents. $a_\mc{L} \in \mc{A}$ denotes the leader agent, while the follower agents are in the set $\mathcal{A} \setminus \{a_\mathcal{L}\}$. Additionally, while it is possible to segment $\mathcal{A}$ into a finite number of leader-follower subsets (e.g., in the multi-leader case \cite{sorensenUnifiedFormationControl2007}), we have assumed that there is only one leader (see assumption (\ref{it:leadfoll})) and that all other agents are followers within any optimization horizon. A few other assumptions will be introduced in later sections as we specify the notation required for their definition. However, with the above setting, we can now present the formation tracking problem as follows: Given $N$ agents in total, $N-1$ followers must keep their positions in the formation while the randomly-selected leader tracks a particular trajectory in space, with possibly inaccurate state information from sensor measurements. Essentially, we require that the group formation be preserved, with the least possible formation error, while the leader agent tracks a specified trajectory. 

    \section{Planar Quadrotor Model}
    \label{sec:2dquad}
\begin{figure}[!t]
    \centering
    \includegraphics[width=0.7\linewidth]{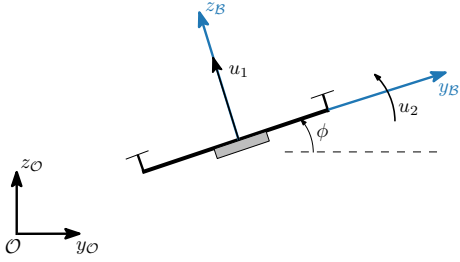}
    \caption{Planar Quadrotor Model and Coordinate Frames: $\mc O$ is the origin of the inertial frame. A moving body frame, subscripted by $\mc B$ and pictured in \textit{blue}, is attached to the agent's center of mass. $u_1$ and $u_2$ retain their former definitions.}
    \label{fig:quadmod}
    \end{figure}

    We model the agents as quadrotors in the plane (see Figure \ref{fig:quadmod}), governed by the dynamics of a planar quadrotor linearized at the equilibrium (hover) state:
    \begin{subequations}
    \label{eq:masmodelss}
    \begin{align}
        &\ddot{y}_i = -g\phi_i\\
        &\ddot{z}_i = -g + \frac{u_{1_i}}{m}\\
        &\ddot{\phi}_i = \frac{u_{2_i}}{I_{xx}},
    \end{align}
    \end{subequations}
    
    where $m$ is the mass of each agent, $g$ is the gravitation constant, and $I_{xx}$ is the $x$ component of the (diagonal) inertia matrix. To simulate sensor noise, we introduce White Gaussian Noise (WGN) terms to the formulation in (\ref{eq:masmodelss}) to get:
    \begin{subequations}
    \label{eq:masmodelssawgn}
    \begin{align}
        &\ddot{y}_i = -g\phi_i + {w}_1\\
        &\ddot{z}_i = -g + \frac{u_{1_i}}{m} + {w}_2\\
        &\ddot{\phi}_i = \frac{u_{2_i}}{I_{xx}} + {w}_3,
    \end{align}

    \end{subequations}
    We can now write the planar quadrotor model in state-space form as:
    
    \begin{equation}
    \label{eq:statewithw}
    \dot{{\mbf x}}_i = f_i({{\mbf x}_i}, {\mbf u}_i, \mbf{w}) = {\mbf A}{\mbf x}_i + {\mbf B}{\mbf u}_i + {\mbf G}_{c}g + {\mbf K}{\mbf w}.
    \end{equation}
    
   where ${\mbf A}$ and ${\mbf B}$ are the plant and input matrices of the state-space model with appropriate dimensions, respectively, with $\text{det}({\mbf A}) \neq 0$. ${\mbf G}_{c}$ is the vector $\begin{bsmallmatrix}0 & 0 & 0 & 0 & -1 & 0 & 0\end{bsmallmatrix}^T$, which accounts for gravity compensation in the $z_\mc{O}$-direction. ${\mbf K} \in \mb{R}^{6\times 6}$ is the noise gain matrix while ${\mbf w} \in \mb{R}^{6}$ is the vector $\begin{bsmallmatrix}0 & w_1 & 0 & w_2 & 0 & w_3\end{bsmallmatrix}^T$, where each $w_j\ (j = [1,2,3]$) follows a Gaussian distribution with mean $\mu \in \mb{R}$ and standard deviation, $\sigma \in \mb{R}$. Table \ref{tab:simquadparam}, partly adapted from \cite{forsterSystemIdentificationCrazyflie2015a}, lists the discussed parameters for the simulated planar quadrotor. 
    
    \begin{table}[htb]
            \caption{Planar Quadrotor Model Parameters}
        \label{tab:simquadparam}
        \centering
        \begin{tabular}{cc}
        \hline
        Parameter & Value\\
         \hline
          \hline
        $m$ [kg] & 0.028\\
        $I_{xx}$ [kg$m^2$] & $6.4893\cdot10^{-6}$\\
        $[\mu \ \sigma ]$ & $[0.0\ 0.2]$ \\
         \hline
        \end{tabular}
    \end{table}

    \section{Optimal Quadrotor Formation Control}
    \label{sec:optconc}

    To achieve the formation tracking task as set forth in Section \ref{sec:probform}, we solve the following initial-value, finite-horizon optimal control problem (FHOCP) for the $i^\text{th}$ agent's control input, ${\mbf u}_i; \ i = [1, 2, \dots, N]$, on the interval $\tau = [0, T]$:

    \begin{subequations}
    \label{eq:optprob}
        \begin{align}
        \min_{{\mbf u}_i} \quad & J_i
        \nonumber\\
        \textrm{subject to:} \quad &  \dot{{\mbf x}}_i(\tau) = f_i({{\mbf x}_i}(\tau),{\mbf u}_i(\tau),{\mbf w})\\ 
        &{{{\mbf{x}}}_i}(0) = {\mbf x}_i^0,\\ \ 
        \label{eq:iadist}
        \quad & \big\lvert\big\lvert {\mbf \Gamma}_i(\tau) -{\mbf \Gamma}_j(\tau)\big\rvert\big\rvert_2 = d_{ij}^r;\ i \neq j \\
        \label{eq:ubound}
        \quad & \lvert {u}_{1_i}\rvert \leq u_{1_{\text{max}}}; \quad \lvert {u}_{2_i}\rvert \leq u_{2_{\text{max}}}.
        \end{align}   
    \end{subequations}

     Here, $J_i$ is the objective for the $\ith$ agent equal to the total expectation of the trajectory error, control, and Mayer costs defined as:

    \begin{equation}
        \label{eq:objexp}
        \mb{E}\Bigg[\int_{\tau = 0}^{T}L_i({{\mbf{\Gamma}}_i}[\tau],{\mbf u}_i[\tau])d\tau + h(\mbf{x}_i(T))\Bigg],
    \end{equation}

    where $L_i({{\mbf{x}}_i}[\tau],{\mbf u}_i[\tau]:\mb{R}^{2}\times\mb{R}^2 \mapsto \mb{R}$ is the Lagrangian defined as follows (the $\tau$ argument has been omitted for brevity): 
    
    \begin{equation}
        \label{eq:lagrgn}
       {\mbf u}_i^T{\mbf R}_i{\mbf u}_i + ({\mbf \Gamma}_i-{{\mbf \Gamma}_i}^r)^T{\mbf Q}_i({\mbf \Gamma}_i-{{\mbf \Gamma}_i}^r),
    \end{equation}
    
    and $h:\mb{R}^{6} \mapsto \mb{R}$ is the terminal (Mayer) cost for the $i^{\text{th}}$ optimal control problem, given as:
    \begin{equation}
       h({{\mbf x}_i}(T)) = {{\mbf x}_i}^T(T){\mbf P}_i{{\mbf x}_i}(T).
    \end{equation}

    In (\ref{eq:objexp}), $\mb{E}$ is the expectation operator --- defined in terms of the instantaneous probabilities $p(s = s(*(\tau))$ --- as:
        \begin{equation}
        \label{eq:expop}
        \mb{E}[s(*)] = \int_{\tau = 0}^T{s(*(\tau))p(s)}d\tau,
    \end{equation}
    
    where $*$ here represents a generic time-dependent argument of $s$, a generic function. In the preceding equations, we denote the set of admissible control laws or policies as $\mc{U} \subseteq \mathbb{R}^2$. $T \in \mb{R}$ is the time horizon for the optimal control problem, and ${{\mbf x}_i} = \begin{bsmallmatrix}
        y_i & \dot{y_i} & z_i & \dot{z_i} & \phi_i & \dot{\phi_i}
    \end{bsmallmatrix}^T \in \mb{R}^{6}$ is the state of the $i^{\text{th}}$ agent. $y_i$ and $z_i$ are the respective positions of the $\ith$ agent along the $Y$ and $Z$ inertial axes, while $\phi_i$ is its roll angle. The ($\dot{\quad}$) variables in $\mbf{x}_i$ represent the corresponding linear ($y$ and $z$) and roll rates. ${{\mbf x}_i}^r\in \mb{R}^{6}$ is the $i^{\text{th}}$ agent's reference state, ${\mbf u}_i = \begin{bsmallmatrix} u_{1_i} & u_{2_i} \end{bsmallmatrix}\in \mc{U}$ is the $i^{\text{th}}$ control, with $u_{1_i}$ and $u_{2_i}$ respectively equal to the effective gravity-opposing force produced by the propellers of the $\ith$ agent and the torque about the suppressed inertial $X$ axis. $\mbf{\Gamma}_i$ is the $\ith$ agent's trajectory equal to the vector $ \begin{bsmallmatrix} y_i & z_i \end{bsmallmatrix}^T \in \mb{R}^{2}$ while $d_{ij}^r \in \mb{R}_+$ is the prescribed inter-agent distance, which can be thought of as representing the limited communication range between agents or as a simple inter-agent proximity constraint for collision avoidance. $\x_i^0$ is the initial state of the $\ith$ agent. ${\mbf P}_i \in \mb{R}^{6\times 6}$ is the weight matrix of the $i^{\text{th}}$ terminal cost, ${\mbf R}_i \in \mb{R}^{2\times 2}$ is the weight matrix corresponding to the control cost, and ${\mbf Q}_i \in \mb{R}^{2\times 2}$ is the weight matrix for the cost corresponding to the $i^{\text{th}}$ trajectory error $({\mbf \Gamma}_i - {{\mbf \Gamma}_i}^r)$, with the $i^{\text{th}}$ time-varying trajectory ${{\mbf \Gamma}_i}^r(\tau)$ as reference. ${\mbf P}_i$ and ${\mbf R}_i$ are taken to be positive definite ($p.d.$) matrices, while ${\mbf Q}_i$ is chosen as follows:

    \begin{equation*} 
            {\mbf Q}_i \ is \ 
                \begin{cases}
                     p.d., & \text{if } a_i = a_\mc{L} \\ 
                     0 \in \mb{R}^{2\times 2}, & \text{otherwise. }
                \end{cases} 
    \end{equation*}

    Since the desired position $\begin{bsmallmatrix} y_{i} & z_{i} \end{bsmallmatrix}$ in the $YZ$-plane encodes the trajectory of the $\ith$ agent, the choice of Lagrangian in (\ref{eq:lagrgn}) -- with ${\mbf Q}_i$ as defined -- ensures that the leader tracks a specific trajectory determined by a high-level trajectory planner (see Figure \ref{fig:optcontarch}), while the other agents maintain their position in the formation. We define this desired formation by specifying rules that guide the inter-agent distances between the leader and follower agents and between the follower agents themselves (see Section \ref{ssec:formgen}). We also set an upper bound on the magnitude of the control signals for each agent ($u_{1_{\text{max}}}$ and $u_{2_{\text{max}}}$), which is standard in practice. $f_i: \mb{R}^{6}\times\mb{R}^2 \mapsto \mb{R}^{6}$ is the continuous linear time-variant state-space model describing the $i^{\text{th}}$ agent, presented in Section \ref{sec:2dquad}. 
    
    With the model in (\ref{eq:statewithw}), we rewrite the optimal control problem (\ref{eq:optprob}) in a more compact fashion as:
    \begin{subequations}
    \label{eq:optprobalt}
        \begin{align}
        \label{eq:newobjfunc}
        \begin{split}
              \min_{{\mbf u}_i} \quad & J_i + \ \lambda_i \cdot (d_{ij}^r - \big\lvert\big\lvert {\mbf \Gamma}_i(t) - {\mbf \Gamma}_j(t)\big\rvert\big\rvert_2) \nonumber
        \end{split}\\
        \textrm{subject to:} \quad &  {\dot{\mbf x}_i}(\tau) = f_i({{\mbf x}_i}(\tau),{\mbf u}_i(\tau),{\mbf w})\\ 
        &{{\mbf x}_i}(0) = {\mbf x}_i^0 \\
        \quad & \lvert {u}_{1_i}\rvert \leq u_{1_{\text{max}}}; \quad \lvert {u}_{2_i}\rvert \leq u_{2_{\text{max}}}, 
        \end{align}
    
    \end{subequations}    
    
 where we have introduced the inter-agent distance constraint as a penalty term in $J_i$.  The objective function in (\ref{eq:optprobalt}) is the $\ith$ augmented Lagrangian. $\lambda_i$ is a non-negative real term that specifies whether the inter-agent distance constraint is taken into account in the $i^\text{th}$ optimal control problem, and to what degree if so. Thus, we set the value for $\lambda_i$ as follows:

            \begin{equation*} 
            \lambda_i = 
                \begin{cases}
                     0, & \text{if } a_i = a_\mc{L} \\ 
                     \beta > 0, & \text{otherwise.}
                \end{cases} 
            \end{equation*} 

With this problem formulation and choice of $\lambda_i$, only the leader tracks the desired trajectory, while the follower agents simply keep their respective positions in the formation as determined by the triangular formation rule and corresponding inter-agent distances.

\section{Formation Specification}
\label{ssec:formgen}

In addition to tight trajectory tracking, we require the MAS to maintain a triangular formation. We elect this formation because it is geometrically well suited to the leader-follower concept and also ensures that the followers are uniformly distributed spatially on a line segment behind the leader. This pattern has been shown to be locally asymptotically stable under the assumption that the formation is infinitesimally rigid \cite{ohFormationControlMobile2011}.

\begin{figure}[H]
		\centering
        \includegraphics[width=0.45\textwidth]{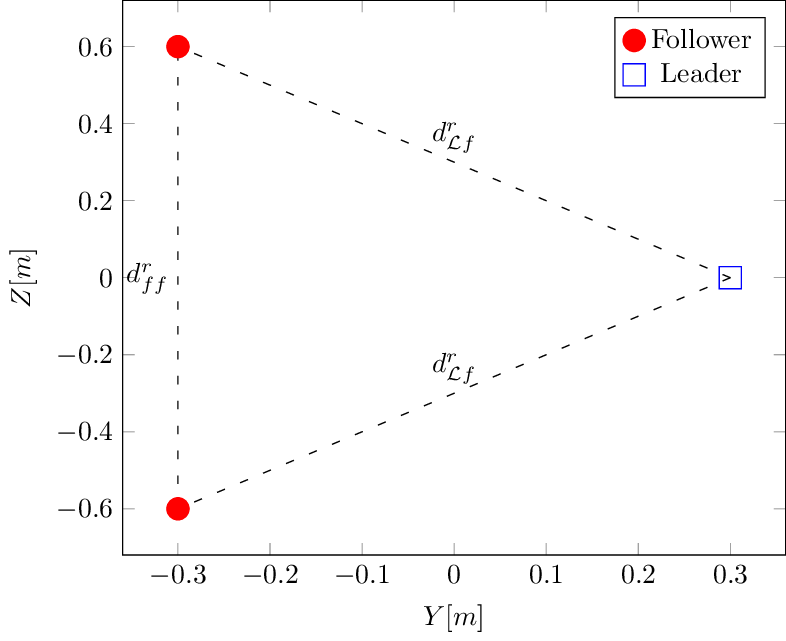}
		\caption{Three agents in the plane assuming a triangular leader-follower formation. $d^r_{\mc{L}f}$ and $d^r_{ff}$ are the desired leader-follower and follower-follower distances, respectively.}
		\label{fig:triform}
	\end{figure}

To this end, we require that the formation be rigid, and translation and rotation invariant, i.e., that:
\begin{equation}
    \big\lvert\big\lvert {\mbf \Gamma}_i - {\mbf \Gamma}_j\big\rvert\big\rvert_2 = d_{ij}^r \ \forall \ i, j \in \{1,2, \dots, N\};\ i\neq j,
\end{equation}
 and that there exists a $\bm{\xi}_i \in \mb{R}^{2}$ for the $\ith$ agent such that:
 \begin{equation}
 \begin{bmatrix}
     \mbf{R}_{\theta} & \bm{\tau}_d\\
      0 & 1
 \end{bmatrix} \cdot 
 \begin{bmatrix}
     \bm{\xi}_i \\
     1
 \end{bmatrix}
 = \begin{bmatrix}
     {\mbf \Gamma}_i \\
     1
 \end{bmatrix},
 \end{equation}

respectively, for some homogeneous transformation in $SE(3)$ comprising a fixed rotation about the $X$ axis by $\theta \in \mb{R}$ -- encoded by the rotation matrix $\mbf{R}_{\theta} \in SO(2)$ -- and a fixed translation by $\bm{\tau}_d \in \mb{R}^2$. Figure \ref{fig:triform} depicts three agents in triangular leader-follower formation with the desired inter-agent distances labeled. 
		
\section{Simulation Studies}
\label{sec:simstudies}
For simulation, we solve (\ref{eq:optprob}) using the Ipopt Python optimization package \cite{wachterImplementationInteriorpointFilter2006}, with the problem setup parameters outlined in Table \ref{tab:params}. $\mathbbm{1}_{n}$ is the $n\times n$ identity matrix. In our problem setup, we also assume that the dynamics of each agent is propagated forward in time.

\begin{table}[htb]
        \caption{FHOCP Setup Parameters}
    \label{tab:params}
    \centering
    \resizebox{\columnwidth}{!}{%
    \begin{tabular}{ccccccccccc}
    \hline
    \multirow{2}{*}{Parameter} & $u_{1_{\text{max}}}$ & $u_{2_{\text{max}}}$ & \multicolumn{2}{c}{${\mbf P}_i$, ${\mbf K}_i$} & \multicolumn{2}{c}{${\mbf Q}_i$, ${\mbf R}_i$} & $d^r_{\mc{L}f}$ & $d^r_{ff}$ & $\beta$ & $T$\\
    & [N] & [N-m] & & & & &[m] & [m]\\
    \hline \hline
     Value  & $1.2mg$  & $\frac{I_{xx}\pi}{10}$ &  \multicolumn{3}{c}{$\mathbbm{1}_{6}$} & $\mathbbm{1}_{2}$ & 0.5 & 0.5 & 1 & 10\\ 
     \hline
    \end{tabular}%
    }
\end{table}

\section{Results \& Analysis}
\label{sec:res}

\subsection{Trajectory Tracking}
\begin{figure}[htb]
    \centering
    \includegraphics[trim=0pt 0pt 0pt 0pt, clip, width=0.8\columnwidth]{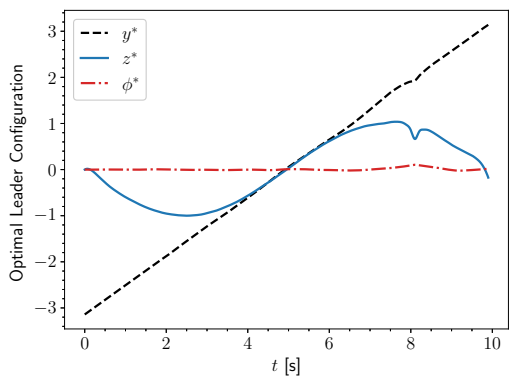}
    \caption{Time evolution of the optimal configuration for the leader agent (model with $\mbf{w}$).}
    \label{fig:allplots}
\end{figure}

\begin{figure}[htb]
        \centering
        \includegraphics[trim=5pt 0pt 0pt 0pt, clip, width=0.8\columnwidth]{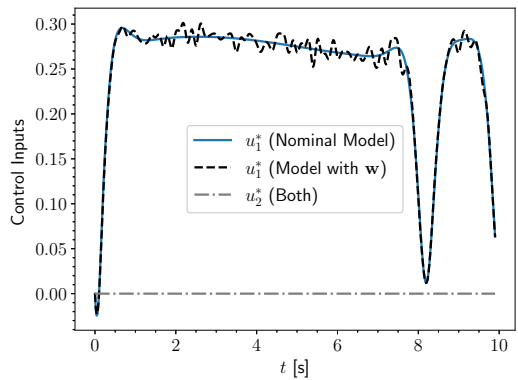}
     \caption{Optimal control inputs for the leader agent.}
    \label{fig:ctsine}
\end{figure}

\begin{figure}[htb]
        \centering
        \includegraphics[trim=5pt 0pt 0pt 0pt, clip, width=0.8\columnwidth]{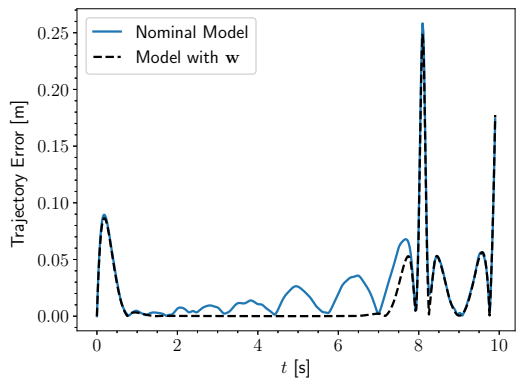}
     \caption{Leader trajectory error.}
    \label{fig:leadtrajerr}
\end{figure}

\begin{figure*}
    \centering
    \begin{subfigure}[b]{0.45\textwidth}
         \centering
         \includegraphics[trim=0pt 0pt 0pt 0pt, clip, width=\sfigsz]{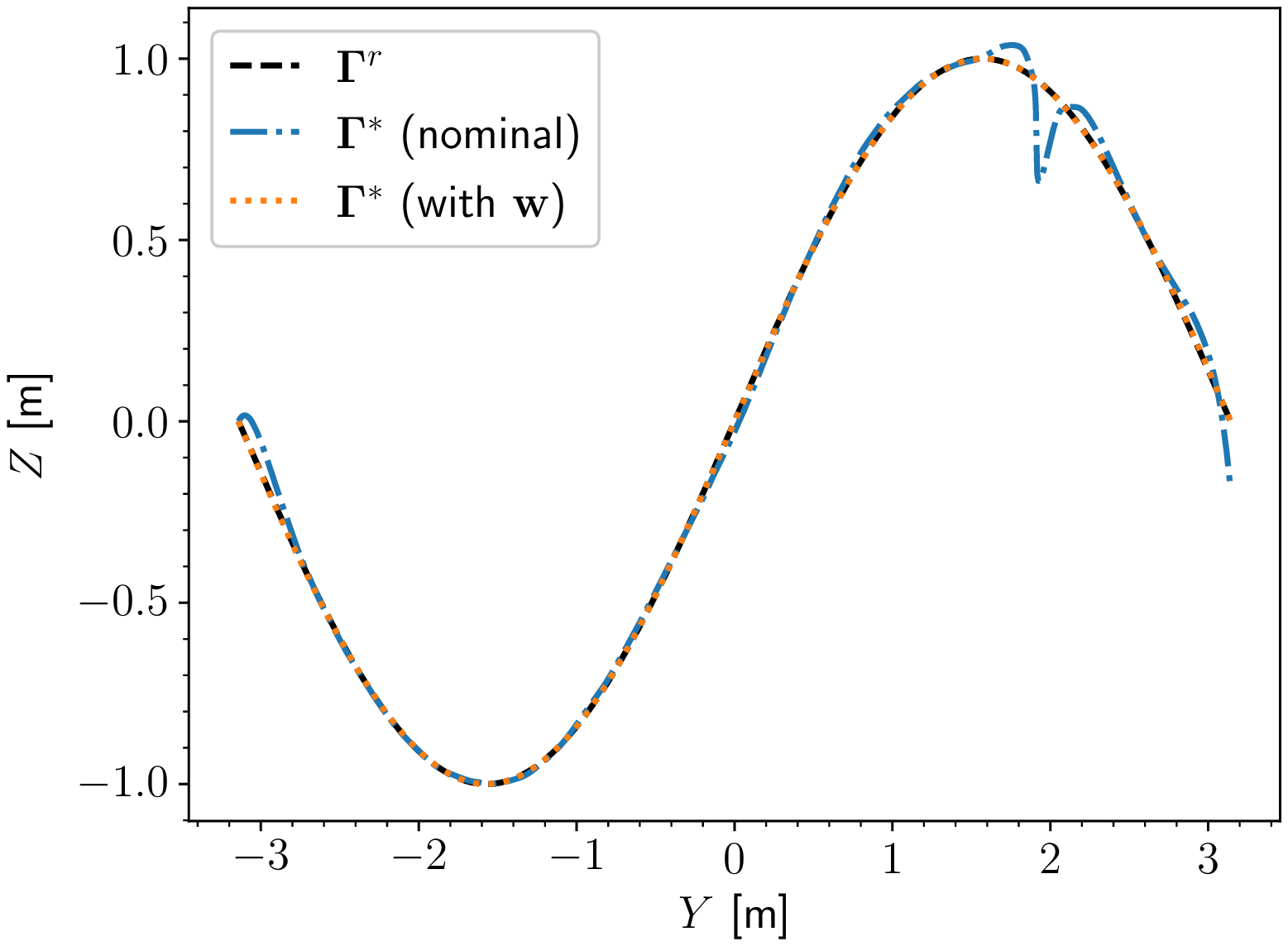}
         \caption{}
         \label{fig:yvsyd}
     \end{subfigure}
     \hfill
     \begin{subfigure}[b]{0.45\textwidth}
         \centering
        \includegraphics[width=\sfigsz]{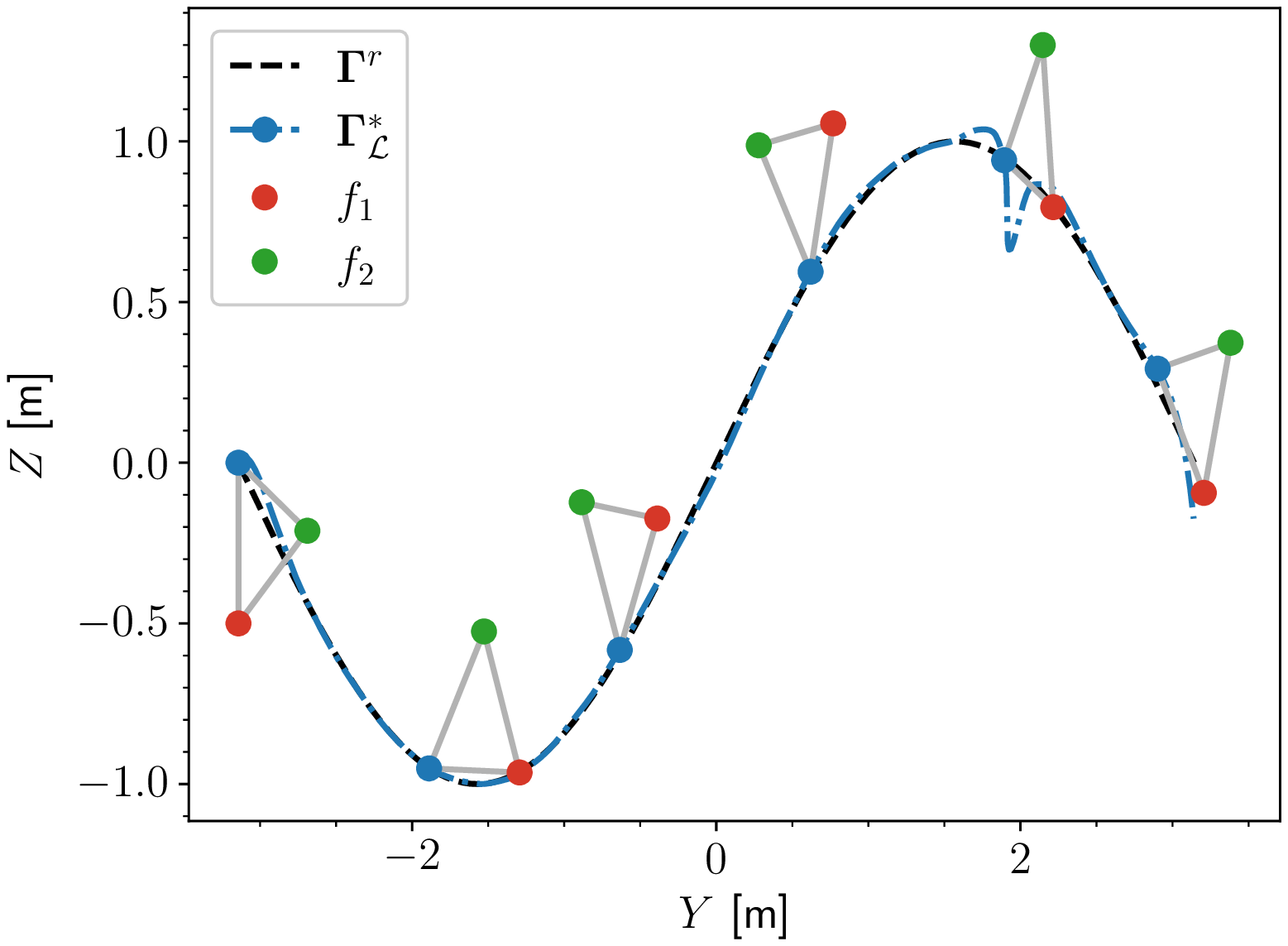}        
        \caption{}
         \label{fig:trackplusform}
     \end{subfigure}
     \caption{(a) Optimal leader agent trajectory with a sinusoidal reference (for both model cases) and (b) A snapshot demonstrating the planar formation tracking performance for the model with $\mbf w$ (over $T$).}
\end{figure*}

Figures \ref{fig:allplots} and \ref{fig:ctsine} show the optimal state variables and control inputs, respectively, for the leader agent. As expected, the optimal $\phi$ and $u_2$ values over the time horizon are approximately zero implying near hover state. Concerning tracking performance, we can see from Figure \ref{fig:yvsyd} that the optimal trajectory closely tracks the desired sinusoidal reference trajectory. As expected, in the case with no measurement noise, a much better tracking performance is recorded -- near zero trajectory error and hence, tight trajectory tracking (Figure \ref{fig:leadtrajerr}). For a numerical comparison, Table \ref{tab:rmse} presents root-mean-square error (RMSE) values for the leader agent's trajectory error, along with those for the error between the desired and actual inter-agent distances for both follower agents (abbreviated as $f_1$ and $f_2$).

\subsection{Triangular Formation Maintenance}
Results that portray the inter-agent distance maintenance results are presented in Figure \ref{fig:trackplusform} and Table \ref{tab:rmse}, from where we observe that, over the time horizon, the triangular formation is maintained as the leader closely tracks the desired trajectory. Upon the introduction of the WGN term, a comparable performance is recorded, with only slight deviations from the trajectory reference and inter-agent distance constraints --- evidenced also by the low RMSE values obtained for all three agents. A sample simulation video (showing the formation tracking performance) is available at the following link: \tee{\url{https://youtu.be/aJJXmN3UJoQ}}.

\begin{table}[htb]
    \caption{RMSE values for the leader and follower agents}
    \label{tab:rmse}
    \centering
    \begin{tabular}{ccc}
    \hline
    \multirow{2}{*}{Agent} & \multicolumn{2}{c}{RMSE}\\ 
    \cline{2-3}
    & Nominal Model & Model with ${\mbf w}$\\
        \hline
        \hline
    \multirow{1}{*}{Leader} &  1.18$\times10^{-7}$ & 0.0836\\

    \multirow{1}{*}{$f_1$} & 1.34$\times 10^{-12}$ &  1.4$\times10^{-9}$\\

    \multirow{1}{*}{$f_2$} & 2.6$\times 10^{-9}$ & 1.3$\times10^{-9}$\\
 
     \hline
    \end{tabular}%
\end{table}

\section{Conclusion \& Future Work}
\label{sec:conc}

This paper presented results from formation tracking experiments for a special class of multiagent systems: quadrotors in the plane, restricted to states near the equilibrium. We showed that for the studied model, an overall acceptable tracking performance was recorded for the MAS, even with simulated white Gaussian noise in the model. In future work, we will apply similar ideas to the case where the agents' states evolve in space under the standard nonlinear quadrotor model. Another promising direction would be to study how the performance of the optimal control scheme is affected by scaling the number of agents.

\section*{Acknowledgements}
The authors gratefully acknowledge Microsoft Corporation and the Maryland Robotics Center for their kind financial support.


\bibliographystyle{ieeetr}
\bibliography{ofc2D}

\end{document}